\documentclass[twocolumn,showpacs,preprintnumbers,amsmath,amssymb]{revtex4}

\usepackage{graphicx}
\usepackage{dcolumn}% Align table columns on decimal point
\usepackage{bm}% bold math

\usepackage{multirow}  
\usepackage{verbatim}  

\begin{document}
\title{Satisfying the Fluctuation Theorem in Free Energy Calculations with
Hamiltonian Replica Exchange } 
\author{Matthew A. Wyczalkowski}
\author{Rohit V. Pappu}
\affiliation{Department of Biomedical Engineering and Center for
Computational Biology\\ 
Washington University in St. Louis, St. Louis, MO 63130 }
%\date{\today}

\begin{abstract}
A novel error measure, referred to as the hysteresis error, is developed from
the Crooks fluctuation theorem to evaluate sampling quality in free energy
calculations.  Theory and numerical free energy of hydration calculations are
used to show that Hamiltonian replica exchange provides a direct route for
minimizing the hysteresis error.  Replica exchange swap probabilies yield the
rate at which the hysteresis error falls with simulation length, and this
result can be used to decrease bias and statistical errors associated with
free energy calculations based on multicanonical simulations.
\end{abstract}

%%%%%%%%%%%%%%%%%%%%%%%%%%%%%%%%%%%%%%%%%%%%%%%%%%%%%%%%%%%%%%%%%%%%%%%%%%%%%%%%%
\maketitle

\section{Introduction and Overview}

Free energies of solvation provide quantitative assessments of driving forces
for spontaneous processes such as protein folding, binding, self-assembly, and
solubility. Formally, the free energy of solvation in the canonical ensemble
is the free energy change $\Delta F$ associated with the transfer of a solute
from the gas phase to a fixed position in the solvent \cite{ben-naim:1987}.
Operationally, one has access to a range of techniques to obtain estimates for
$\Delta F$ \cite{ytreberg:2006,levy:1998}.  Kirkwood \cite{kirkwood:1935}
showed that one could introduce arbitrary parameters into potential functions
and continuously vary the degree of coupling between specific molecules in a
dense fluid. The device of coupling parameters leads to simple expressions for
chemical potentials of any component of the fluid.  If the component is the
solute molecule, which is transferred from the gas phase into the solvent,
then a single coupling parameter $\lambda$, where $0 \leq \lambda \leq 1$,
modulates solute-solvent interactions in the system's potential function. The
limits $\lambda=0$ and $\lambda=1$ correspond to the pure solvent and solvent
plus fully grown solute, respectively. Intermediate values of $\lambda$
correspond to potential functions that include only a part of the
solute-solvent interactions. The Kirkwood coupling parameter plays a central
role in equilibrium methods for calculating $\Delta F$.  One carries out a
series of independent canonical simulations where each simulation is
associated with a distinct potential function,  characterized by  a specific
$\lambda$ value.  As it samples the equilibrium ensemble, each simulation
generates a series of work values, which are then used to estimate the free
energy change across the entire $\lambda$ schedule.

The multicanonical approach described above takes advantage of the simple
formalism developed by Kirkwood for calculating $\Delta F$. However, in
practice,  standard free energy calculations based on multicanonical
simulations are plagued by slow convergence and inaccurate estimates of
$\Delta F$ \cite{mobley:2007}.  Errors may be divided into statistical and
bias (or finite sampling) errors \cite{lu:2001a}.  The former stem from the
fluctuations of
the free energy estimator, and can be estimated by block averaging or
bootstrap methods \cite{frenkel:2002,efron:1993}.  Since the statistical error
decreases as the inverse square root of simulation length, it is frequently
used as an indicator of the convergence of the multicanonical simulation.
While statistical errors are random fluctuations of short simulation results
about some mean value, the bias error is an error of the mean value itself,
and it changes with simulation length.  As discussed by
Zuckerman and Woolf\cite{zuckerman:2004}, bias errors have two causes: the free energy
estimates are nonlinear averages; and, the work distributions on which such
estimates are based will typically have long tails which are rarely sampled,
and yet these are important to the average.  The latter point is important:
rare events dominate free energy estimates, and one seldom observes these
events in short simulations. As a result, the average drifts with simulation
length, resulting in inaccurate estimates for $\Delta F$ from bias error even
when the statistical error is small.  The magnitude of the bias error is
difficult to quantify directly, as it requires knowledge of the actual free
energy difference, the very quantity we wish to determine. Furthermore, small
fluctuations in the estimate for $\Delta F$ may not be indicative of
convergence, but rather of inadequate sampling of the rare but important
configurations.  To address these problems, we develop an alternate measure of
free energy error, one based on deviations from equilibrium distributions.

Crooks \cite{crooks:1999} derived a fluctuation theorem (appendix \ref{sec:ft})
valid for stochastic, microscopically reversible dynamics, which relates the
distribution of dissipated work values along a forward and reverse path as,
\begin{equation}
\exp(\beta W_D) = \frac{P_F(\beta W_D)}{P_R(-\beta W_D)} \label{eqn:Pwork}.
\end{equation}
Here, $\beta=(k_BT)^{-1}$, $P_F(\beta W_D)$ is the probability distribution
for dissipated work associated with switching $\lambda$ from $\lambda_0$ to
$\lambda_1$, and $P_R(-\beta W_D)$ is the corresponding distribution for the
reverse process.  If the canonical simulations for each value of $\lambda$
sample the equilibrium ensemble adequately, then the distributions of
dissipated work obtained over the course of free energy calculations will
satisfy Eq.  \eqref{eqn:Pwork}.

In this work, we develop a readily measured error estimate, the hysteresis
error $\epsilon_H$, which quantifies the degree to which observed work
distributions obey the Crooks fluctuation theorem.  Hamiltonian replica
exchange, a multicanonical equilibration technique, effectively reduces the
hysteresis error.  We relate the average replica exchange swap probability to
the degree of overlap between equilibrium ensembles, as well as to the rate at
which $\epsilon_H$ falls.  Based on this, we may construct an optimized
$\lambda$ schedule to further minimize the hysteresis error for an entire
simulation. 

The remainder of this presentation is organized as follows: the theory section
introduces the hysteresis error in the context of the Crooks fluctuation
theorem followed by a formal illustration of how Hamiltonian replica exchange minimizes
$\epsilon_H$; the definition of swap probability as a measure of the overlap
between different equilibrium ensembles; and a connection between the amount
of overlap and minimization of $\epsilon_H$. We calculate the free energy of
hydration for acetamide to demonstrate how to estimate $\epsilon_H$ and
minimize this error using replica exchange coupled to standard multicanonical
simulations. We conclude with a summary and a discussion of the features of
our methodology.

%%%%%%%%%%%%%%%%%%%%%%%%%%%%%%%%%%%%%%%%%%%%%%%%%%%%%%%%%%%%%%%%%%%%%%%%%%%%%%%
\section{Theory}

\subsection{Background}
The free energy of replica $i$ in the canonical ensemble at temperature $T$,
whose potential $U_i(\Gamma) = U(\Gamma, \lambda_i)$ is a function of system
configuration $\Gamma$ and the parameter $\lambda_i$, is formally given
as \cite{chandler:1987},
\begin{equation}
F_i = \beta^{-1} \ln\left\{\int d\Gamma \exp[-\beta U_i(\Gamma)]\right \}.
\end{equation}
At equilibrium, the probability of observing configuration $\Gamma$ is given as,
\begin{equation}
\rho_i(\Gamma) = \exp\{\beta [F_i - U_i(\Gamma)]\} \label{eqn:rho_FU}.
\end{equation}

To calculate the free energy change $\delta F$ associated with switching the
Hamiltonian from $U_0$ to $U_1$ we perform simulations at $\lambda_0$ and
$\lambda_1$, and calculate the forward and reverse work as,
\begin{subequations} 
\label{eqn:work}   
\begin{align}
W^F(\Gamma) &= U_1(\Gamma) - U_0(\Gamma) \label{eqn:wf},\\
W^R(\Gamma) &= U_0(\Gamma) - U_1(\Gamma) \label{eqn:wr}.
\end{align}
\end{subequations}
For the forward and reverse work values the configuration $\Gamma$ is
typically drawn from the equilibrium ensemble of $U_0$ and $U_1$,
respectively.  The Free Energy Perturbation (FEP) method \cite{zwanzig:1954}
utilizes forward and reverse work distributions to provide two independent
estimators for $\delta F$,
\begin{subequations} 
\label{eqn:fep}
\begin{eqnarray}
\delta F_{FEP}^F &=& 
   -\beta^{-1} \ln \langle \exp(-\beta W^F) \rangle_0 \label{eqn:fep_F}, \\
\delta F_{FEP}^R  &=&  
   +\beta^{-1} \ln \langle \exp(-\beta W^R) \rangle_1 \label{eqn:fep_R},
\end{eqnarray}
\end{subequations} 
where the forward estimator $\delta F_{FEP}^F$ utilizes forward work values
from the simulation at $U_0$, and the reverse estimator the reverse work from
$U_1$.  Note that in both cases $\delta F$ is associated with the process of
switching $\lambda_0 \rightarrow \lambda_1$.  These two estimators have
different convergence rates \cite{lu:2001a}. Therefore, while in practice the
two estimates should be equal, in simulations with finite sampling they are
generally different.

Another free energy estimator, the Bennett Acceptance
Ratio \cite{bennett:1976}, uses both the $W^F$ and $W^R$ distributions to
obtain a free energy estimate.  It is generally more
accurate \cite{shirts:2005b} and is employed later in this paper for numerical
free energy estimates, but will not be considered for theoretical development.

\subsection{The Hysteresis Error}
% see notes 10/25/07.1 for review of bias, statistical errors.
The hysteresis error $\epsilon_H$ is defined as the difference between the
forward and reverse $\delta F_{FEP}$ estimates,
\begin{equation}
\epsilon_H \equiv \delta F_{FEP}^F - \delta F_{FEP}^R \label{eqn:hyst_def}.
\end{equation}
$\epsilon_H$ has contributions from both the statistical and bias error of the
FEP estimators \cite{lu:2001a,zuckerman:2004}.  The bias 
error of the two estimators is typically in the opposite direction.  While the
statistical error may dominate the $\epsilon_H$ for a given simulation, in
averages over multiple short simulations the dominant contribution to the
average hysteresis error is the sum of the forward and reverse FEP bias.

We take $\epsilon_H$ as a measure of sampling quality and aim to minimize its
magnitude between all pairs of neighboring replicas.  The validity of using
$\epsilon_H$ as a general sampling error is based on a relationship between it
and the fluctuation theorem of Crooks\eqref{eqn:Pwork}, derived below.

Switching the parameter $\lambda_0 \rightarrow \lambda_1$ (and vice versa) is
equivalent to performing non-equilibrium work; the difference between the work
performed and the free energy change of the system is the dissipated work,
defined in the forward and reverse direction as,
\begin{subequations} 
\label{eqn:work_d}   
\begin{align}
W_D^F(\Gamma) &= W^F(\Gamma) - \delta F \label{eqn:wdf},\\
W_D^R(\Gamma) &= W^R(\Gamma) + \delta F \label{eqn:wdr}.
\end{align}
\end{subequations}
Crooks \cite{crooks:1999} equates $W_D^F$ and $W_D^R$ to the entropy production
caused by changing $\lambda_0 \rightarrow \lambda_1$ and $\lambda_1
\rightarrow \lambda_0$, respectively, for the given configuration.

The distributions $P_F(W_D)$ and $P_R(W_D)$ give the probability of realizing
a specific value for the dissipated work in the forward and reverse
directions, respectively. The distributions are related to each other by the
fluctuation theorem shown in Eq. \eqref{eqn:Pwork}, which we have re-derived
in appendix \ref{sec:ft} for the specific case of instantaneous switching
between configurations with different $\lambda$ values.  In practice, Eq.
\eqref{eqn:Pwork} will not be satisfied exactly because of errors due to
finite sampling.  To take simulation errors into account, we rewrite Eq.
\eqref{eqn:Pwork} with an arbitrary error term $\epsilon_{FT}^\ast$ and with
observed (rather than ideal) dissipated work distributions $P_F^\ast$ and
$P_R^\ast$, 
\begin{equation}
\exp[\beta W_D + \beta \epsilon_{FT}^\ast(W_D)] = \frac{P_F^\ast(\beta
W_D)}{P_R^\ast(-\beta W_D)}. \label{eqn:FT_error}
\end{equation}
Eq. \eqref{eqn:FT_error} is constructed such that the Crooks fluctuation
theorem is recovered and $\epsilon_{FT}^\ast = 0$ when the observed work
distributions match the correct distributions.  The hysteresis error 
$\epsilon_H$ and the fluctuation error $\epsilon^\ast_{FT}$ are related to
each other as, (see appendix \ref{sec:hysteresis_error}),% see notes 11/2/07
\begin{equation}
\epsilon_H = -\beta^{-1} \ln \langle \exp (-\beta
\epsilon^\ast_{FT})\rangle^\ast_0,
\label{eqn:hyst_error_FT} 
\end{equation}
where $\langle \cdot \rangle^\ast$ is defined as the average obtained from a
finite simulation.  The more closely a simulation obeys the relationship
\eqref{eqn:Pwork}, the smaller the hysteresis error $\epsilon_H$, and vice
versa.  In the next section, we will discuss methods to reduce $\epsilon_H$,
which in turn leads to the satisfaction of the Crooks fluctuation theorem.

\subsection{Replica Exchange}
In a Hamiltonian replica exchange \cite{sugita:1999, fukunishi:2002}
simulation, Monte Carlo moves are employed to exchange configurations $\Gamma$
(or equivalently, parameters $\lambda$) between two replicas with the
probability, 
\begin{equation}
P_{swap} = \min[1, \exp(-\beta \Delta U_{swap})] \label{eqn:Pswap_def},
\end{equation}
where,
\begin{subequations}
\begin{eqnarray}
\Delta U_{swap} &=& U_0(\Gamma_1) + U_1(\Gamma_0) \nonumber \\
&&- U_0(\Gamma_0) - U_1(\Gamma_1)\label{eqn:deltaU_def},\\
& = & W^F + W^R \label{eqn:deltaU_W}, \\
& = & W^F_D + W^R_D \label{eqn:deltaU_Wd}.
\end{eqnarray}
\end{subequations}
$\Gamma_0$ and $\Gamma_1$ denote configurations drawn at random from the
equilibrium ensembles of $U_0$ and $U_1$, respectively.  For convenience, we
write $\gamma = (\Gamma_0, \Gamma_1)$ as a pair of such configurations, and
$\gamma' = (\Gamma_1, \Gamma_0)$ is the swapped configuration pair.

\begin{figure*}
\begin{center}
\includegraphics[width=6.25in]{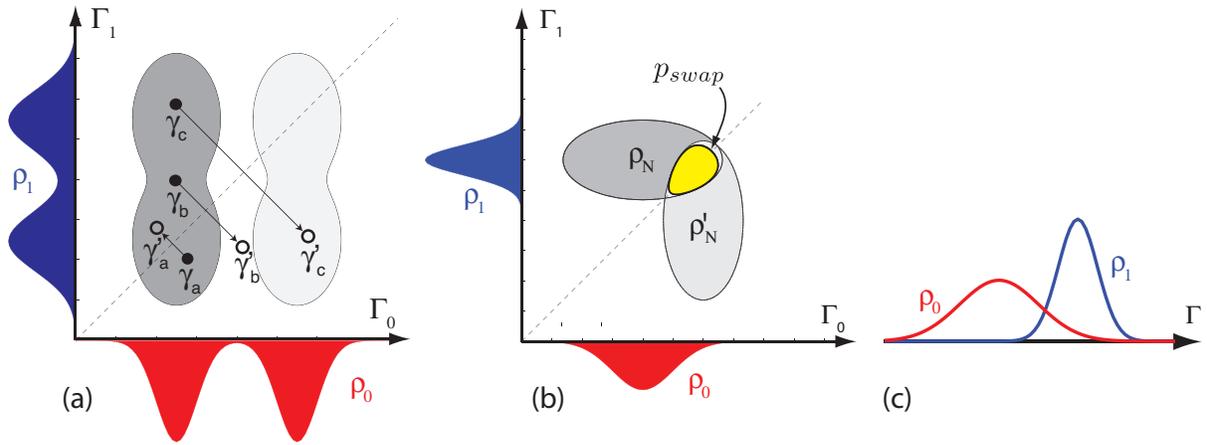}
\end{center}
\caption{
A graphical representation of replica exchange.  {\bf(a)} The independent
(high dimensional) configuration spaces $\Gamma_0$ and $\Gamma_1$ have
probability distributions $\rho_0$ and $\rho_1$, respectively, and the joint
equilibrium ensemble $\rho_N$ is drawn over this domain.  The $\rho_0$ system
has a kinetic barrier (represented by the two disconnected lobes) and with no
replica exchange the system explores only the configurations of the shaded
domain.  A replica exchange swap is a reflection of the configuration pair
$\gamma$ about the $\Gamma_0 = \Gamma_1$ diagonal axis, and three swap
attempts are shown: the configuration pair $\gamma_a$ swaps successfully and
becomes $\gamma'_a$, but it does not sample otherwise inaccessible regions; a
swap of $\gamma_b$ fails because $\gamma'_b$ is not in the equilibrium
ensemble; and the swap of $\gamma_c$ succeeds and allows the system to explore
otherwise inaccessible regions of phase space.  {\bf (b)} The equilibrium
domain $\rho_N$ and its swapped image $\rho'_N$ are drawn.  Swaps are feasible
only for configuration pairs which belong to both $\rho_N$ and $\rho'_N$. This
overlap region, labeled $p_{swap}$, is the domain where the integrand of Eq.
\eqref{eqn:pswap_int} is large, and its size corresponds to the average
swap probability.  {\bf (c)} The overlap of the $\rho_0$ and $\rho_1$
distributions along the common configuration $\Gamma_0 = \Gamma_1$.  For the
hysteresis error to converge, the $\lambda_0$ simulation must observe
configurations where $\rho_1 > \rho_0$, and the $\lambda_1$ simulation must
adequately sample the region $\rho_0 > \rho_1$. The frequency with which this
occurs is given by $\langle p_{swap} \rangle$.}
\label{fig:swap_probability} 
\end{figure*}

Since $\Gamma_0$ and $\Gamma_1$ are independent configurations, we can
consider the probability of sampling $\Gamma_0$ in the equilibrium ensemble of
$U_0$ {\em and} sampling $\Gamma_1$ in the equilibrium ensemble of $U_1$; this
is the native probability $\rho_N(\gamma)$. Analogously, the joint probability of
sampling the swapped configurations, $\Gamma_1$ from $\rho_0$ and $\Gamma_0$
from $\rho_1$ is given as $\rho'_N(\gamma)$:
\begin{subequations}
\label{eqn:rhoNS}
\begin{eqnarray}
\rho_N(\gamma) &=& \rho_0(\Gamma_0) \rho_1(\Gamma_1), \\
\rho'_N(\gamma) &=& \rho_0(\Gamma_1) \rho_1(\Gamma_0) =
\rho_N(\gamma')\label{eqn:rnp_def}.
\end{eqnarray}
\end{subequations}
Replica exchange swaps are conveniently visualized by plotting the
independent configurations $\Gamma_0$ and $\Gamma_1$ along orthogonal axes and
the equilibrium ensemble of the system as an isocontour of $\rho_N$, 
illustrated in Fig.  \ref{fig:swap_probability}(a).  

At equilibrium, the relative probability of observing a pair of replicas in
their swapped versus native configurations is,
\begin{equation}
\frac{\rho'_N}{\rho_N} = \exp(-\beta \Delta U_{swap}), \label{eqn:db}
\end{equation}
which is derived with definitions \eqref{eqn:rhoNS}, \eqref{eqn:rho_FU} and
\eqref{eqn:deltaU_def}.  We will refer to this as an inter-replica equilibrium
relationship.

In an infinitely long simulation, \eqref{eqn:db} will be satisfied exactly, but
this will generally not be the case for finite simulations, where inadequate
sampling of configuration space will result in inaccurate probability
estimates.  However, in simulations with replica exchange we expect the
inter-replica equilibrium relationship to be satisfied more closely than in
simulations without replica exchange, because the swap move distributes
configuration pairs in such a way as to satisfy Eq. \eqref{eqn:db}.  To
illustrate, consider the system in Fig. \ref{fig:swap_probability}(a) where
the $U_0$ replica is presumed to be stuck in the left lobe of the $\rho_0$
distribution because of a kinetic barrier.  Without replica exchange, only the
shaded region of $\rho_N$ will be sampled accurately.  The simulation will not
have a correct estimate for $\rho'_N(\gamma_c) = \rho_N(\gamma'_c)$, since
$\rho_0$ for the swapped configuration, never having been observed, will be
inaccurate.  Consequently, Eq. \eqref{eqn:db} will not hold.  Replica exchange
directly populates swapped configurations (e.g., $\gamma'_c$), thereby
improving the statistics of $\rho'_N$ and allowing inter-replica equilibrium
to be achieved more quickly for all configurations in $\rho_N$.

The degree to which Eq. \eqref{eqn:db} is satisfied determines the magnitude
of the hysteresis error.  To illustrate this, suppose that the distribution
$\rho'_N$ has some small error $\rho_\epsilon(\Gamma_0, \Gamma_1)$ due to
finite sampling, so that we write $(\rho'_N + \rho_\epsilon)$ as the numerator
in Eq. \eqref{eqn:db}.  In appendix \ref{sec:rho_e} we show, by integrating over
all configuration pairs, that the relationship between the hysteresis error
and the error of sampling the swapped distribution, $\rho_\epsilon$ is, 
\begin{equation}
\epsilon_H \simeq -\beta^{-1} \int d\Gamma_0 d\Gamma_1 \rho_\epsilon 
\label{eqn:epsilon_h_rho_S}.
\end{equation}
The hysteresis error, then, will be minimized when the estimated swapped
configuration probabilities $\rho'_N$ are consistent with the equilibrium
distribution.  Since replica exchange populates the swapped configurations
directly, it provides an efficient route to minimizing $\epsilon_H$.  

\subsection{Swap Probability}
Analysis of the average swap probability is complicated by the fact that the
Metropolis function (Eq. \eqref{eqn:Pswap_def}) is not analytical.  For the
purposes of interpreting this quantity, we will instead consider the Fermi
swap probability, 
$$p_{swap} = f(\beta \Delta U_{swap}),$$
where $f(x)$ is defined as,
\begin{equation}
f(x) = 1 / [1 + \exp(x)]. \label{eqn:fermi}
\end{equation}
(See \cite{bennett:1976} for discussion).  We use $p_{swap}$ to denote the
Fermi swap probability and $P_{swap}$ for the Metropolis swap probability;
while the theoretical development uses $p_{swap}$, replica exchange moves are
accepted/rejected using $P_{swap}$.  A simulation with either the Metropolis
or Fermi swap probability will yield a Boltzmann distribution of swapped and
unswapped configurations (Eq. \eqref{eqn:db}).  While the exact numerical
values of the Fermi and Metropolis swap probabilities will differ somewhat,
their qualitative behavior and the conclusions drawn here will hold for both.  

The average Fermi swap probability for two systems evolving independently is,
\begin{subequations}
\begin{eqnarray}
\langle p_{swap} \rangle &\equiv&
\langle \langle f(\beta \Delta U_{swap})  \rangle_0 \rangle_1,
\label{eqn:pswap_def}\\
&=& \int d\Gamma_0 d\Gamma_1 \rho_N f(\beta \Delta U_{swap}),
\end{eqnarray}
\end{subequations}
which can be written as,
\begin{subequations}
\begin{eqnarray}
\langle p_{swap} \rangle &=& \left \langle \left \langle \frac
{\rho'_N}{\rho_N + \rho'_N} \right
\rangle_0 \right \rangle_1,\label{eqn:pswap_ens}\\
&=&  \int d\Gamma_0 d\Gamma_1 \frac {\rho_N \rho'_N}{\rho_N + \rho'_N}.
\label{eqn:pswap_int}
\end{eqnarray}
\end{subequations}
The integrand of \eqref{eqn:pswap_int} is a normalized probability of
observing a given configuration pair, and the average swap probability is then
the overlap of $\rho_N$ and $\rho'_N$.  See Fig. \ref{fig:swap_probability}(b)
for a graphical interpretation.  Thus, a large average swap probability
implies a large overlap between the equilibrium distributions of the two
replicas, and a low $\langle p_{swap}\rangle$ indicates that the
configurations these replicas adopt are distinct.  

We can expand \eqref{eqn:pswap_def} in a Taylor series
about $\lambda = \lambda_0 + \delta_\lambda$.  To leading order in
$\delta_\lambda$, we find that in the neighborhood of $\lambda_0$ the average
swap probability is, (see appendix \ref{sec:pswap}), %cf notes 3/23/07.3
\begin{equation}
\langle p_{swap}\rangle \simeq \frac{1}{2} - 
\frac{\beta^2 \delta_\lambda^2}{4} C_{\lambda},
\label{eqn:pswap_lin}
\end{equation}
where
$$ C_\lambda \equiv \mbox{var}\left( \frac{\partial U}{\partial \lambda}
\right) = 
\left \langle \left (\partial U / \partial \lambda\right)^2 \right \rangle_0 - 
\Bigl  \langle \partial U / \partial \lambda \Bigr \rangle^2_0. $$ 
$C_\lambda$, then, determines the rate at which the average swap probability
declines as the difference in $\lambda$ between the two replicas,
$\delta_\lambda$, increases, although this linear analysis is accurate only
for small $\delta_\lambda$.

\subsection{Swap Probability and the Hysteresis Error Convergence Rate}
We now demonstrate that the average swap probability between two replicas
gives a measure of  how quickly the hysteresis error decreases, on average,
over the course of a simulation.  The hysteresis error is the difference
between the forward and reverse $\delta F_{FEP}$, and since the forward and
reverse FEP estimators do not converge at equal rates \cite{lu:2001a}, it is
the slower of these which governs the convergence of $\epsilon_H$.

We may rewrite Eq. \eqref{eqn:fep_F} as,
\begin{equation}
\langle \exp(-\beta W_D^F) \rangle_0 = 1\label{eqn:WDF_converge}.
\end{equation}
For this to hold, we must sample configurations where $W_D^F<0$; since
the dissipated work is on average greater than zero by the second law of
thermodynamics, such configurations tend to be rare \cite{jarzynski:2006}.  As
a result, the convergence rate of $\delta F_{FEP}^F$ is governed by the
probability of observing negative dissipated forward work values.  Likewise, the
convergence of $\delta F_{FEP}^R$ is dictated by observations of $W_D^R<0$.
We can understand this criterion graphically with the relationships, (see
appendix \ref{sec:ft}),
\begin{subequations}
\begin{eqnarray}
\frac{\rho_0(\Gamma_0)}{\rho_1(\Gamma_0)} &= \exp[\beta W_D^F(\Gamma_0)] 
\label{eqn:wdf_rho},\\
\frac{\rho_1(\Gamma_1)}{\rho_0(\Gamma_1)} &= \exp[\beta W_D^R(\Gamma_1)] 
\label{eqn:wdr_rho}.
\end{eqnarray}
\end{subequations}
In the context of Fig. \ref{fig:swap_probability}(c), observing
$W_D^F<0$ corresponds to sampling configurations from the $\rho_0$
distribution where $\rho_1 > \rho_0$, and for $W_D^R<0$ we require $\rho_0 >
\rho_1$ when sampled from the $\rho_1$ distribution.

Turning our attention to the average swap probability, we note that $\Delta
U_{swap}$, which is the sum of $W_D^F$ and $W_D^R$, is negative whenever
$\rho'_N > \rho_N$ (by Eq. \eqref{eqn:db}).  Configurations for which
this is the case are sampled by a simulation only in the lower-right half of
the domain labeled $p_{swap}$ in Fig. \ref{fig:swap_probability}(b).  The
larger this domain, whose size is given by the average swap probability, the
more frequently negative values of $W_D^F$ and $W_D^R$ are observed,
and the more quickly the hysteresis error converges.  A numerical confirmation
of this argument, that low swap probabilities correspond to large hysteresis
errors and vice versa, is demonstrated in the results section.

%%%%%%%%%%%%%%%%%%%%%%%%%%%%%%%%%%%%%%%%%%%%%%%%%%%%%%%%%%%%%%%%%%%%%%%%%%%%%%%
\section{Methods}

The computational system consists of 21 replicas, each with a different
$\lambda$, which are simulated independently to obtain equilibrium statistics.
The parameter $\lambda$ controls the non-bonded interactions between an
acetamide (ACE) solute and the water molecules. Two independent sets of
simulations were performed, with and without replica exchange, in order to
investigate the effect of this technique.

The Lennard-Jones and Coulomb interactions between the water and ACE molecules
are scaled by $\lambda_{LJ}$ and $\lambda_C$, respectively.  We scaled both
parameters simultaneously, such that $\lambda_{LJ} = \lambda_C$; the single
parameter $\lambda$ then refers to both terms.  The specific way in which the
Lennard-Jones and Coulomb terms scale with $\lambda$ is described in appendix
\ref{sec:scaling}.  $\lambda$ varies across the 21 replicas from 0 to 1 in
increments of 0.05.

Each replica consists of 343 water molecules and one ACE molecule, which is
rigid and whose position is fixed in the central box.  All simulations were
performed at constant temperature (298K) and volume (21.8\AA{} cubic box)
using Metropolis Monte Carlo sampling.  Parameters from the OPLS-AA force
field \cite{jorgensen:1996} and 4-site TIP4P water model \cite{jorgensen:1983}
were used to model the solute and solvent, respectively.  Minimum image
boundary conditions and spherical cutoffs were employed for the Coulomb and
Lennard-Jones potentials.  The cutoff radius was 10.5\AA{} for electrostatic
interactions and 10\AA{} for van der Waals interactions.  Cutoffs were
group-based for the former, and atom-based for the latter.  No long-range
corrections were employed.  All simulations were carried out using the MCCCS
Towhee \cite{towhee} Monte Carlo simulation package\footnote{See
http://towhee.sourceforge.net.  Version maw4\_17\_4 (maw-dev branch) was used
for this work.}.  

The initial configurations for all replicas were identical and correspond to
the end-point of a pre-equilibration run with ACE in water.  For each replica,
simulations consisted of 2 million cycles, where a cycle corresponds to 343
Monte Carlo moves; each move combines rotations and translations of a randomly
chosen individual water molecule.  The initial $10^5$ cycles were discarded
for equilibration.  The average acceptance rate for all replicas was 31\%.  

The replica exchange simulation consists of a number of simulation rounds,
where each replica evolves independently, separated by swap rounds, when a
number of swap attempts take place.  The length of the simulation round was
drawn from a normal distribution with a mean of 500 and standard deviation of
50 cycles.  500 cycles is the approximate energy autocorrelation ``time''.
The swap round consists of $21^2$ swap attempts between randomly selected
replica pairs.  Allowing swaps beyond neighboring replicas increases the
efficiency of replica exchange,  by allowing a replica to traverse the entire
range of $\lambda$ from 0 to 1 more quickly than if only neighbor swaps were
permitted \cite{predescu:2005}.  

During the course of the simulation, the native ($U_i(\Gamma_i)$) and foreign
($U_{j\ne i}(\Gamma_i)$) potential energies, as well as values for
$dU/d\lambda_C$ and $dU/d\lambda_{LJ}$ (where $dU/d\lambda = dU/d\lambda_{LJ}
+ dU/d\lambda_C$), were saved every 10 cycles.  These were then post-processed
to obtain the free energies, the hysteresis error, swap probabilities, and
$C_\lambda$, regardless of whether actual replica exchange swaps took place.
The total free energy of hydration, $\Delta F$, is the sum of all free energy
changes $(\delta F)_i$ between neighboring replicas $i$ and $i+1$, calculated
using the Bennett Acceptance Ratio method \cite{bennett:1976},  
\begin{equation*}
\Delta F \equiv \sum_i^{M-1} (\delta F)_i
\end{equation*}
where $M$ is the total number of replicas.  Similarly, the RMS hysteresis error
$\epsilon_{RMS}$ is the root-mean-square of the hysteresis error
$(\epsilon_H)_i$ between neighboring replicas,
\begin{equation*}
\epsilon_{RMS} \equiv \sqrt{\sum_i^{M-1} (\epsilon_H)_i^2 / M}
\end{equation*}

Statistical errors for $\Delta F$ were estimated using the bootstrap
method \cite{efron:1993}.  With the simulation dataset consisting of N
observations, we drew $n^\ast$ observations at random and with replacement to
create one bootstrap estimate, $\Delta F^\ast$.  This process was repeated
10,000 times, and the standard deviation among all the $\Delta F^\ast$ is the
estimated error of $\Delta F$.  $n^\ast$ is the expected number of independent
observations in the dataset; here, $n^\ast=1900$ with the assumption that
there is one independent observation per two internal energy autocorrelation
``times'' \cite{newman:1999}.

%%%%%%%%%%%%%%%%%%%%%%%%%%%%%%%%%%%%%%%%%%%%%%%%%%%%%%%%%%%%%%%%%%%%%%%%%%%%%%%
\section{Results}

\subsection{Acetamide Free Energy of Hydration}
The hydration free energies we calculate for acetamide are in line with
results obtained by other researchers, as shown in Table \ref{tab:ACE}.  All
numerical results differ somewhat from experimental values due to differences
in force field parameters.  Our calculations were carried out in the canonical
ensemble.  Therefore, we obtain estimates for the Helmholtz free energy
$\Delta F$, whereas the experimental and other computational values obtain
estimates for the Gibbs free energy, $\Delta G$.  However, the distinction
between these two values should be negligible \cite{qian:1996}.  The
consistency between our results and those of others serves to verify our
implementation and sampling technique.  

Table \ref{tab:ACE} shows differences between results obtained with and
without replica exchange.  As expected from our theoretical considerations, we
find that the RMS hysteresis error is lowered by an order of magnitude when
replica exchange is coupled to the multicanonical sampling protocol.  However, it should be
noted that the statistical error estimated using bootstrap remains unaffected.
This is not an artifact of the bootstrap method used to estimate statistical
errors.  Instead, fluctuations in estimates for $\delta F$ originate in
fluctuations of the underlying work distribution, shown in Eq.
\eqref{eqn:Pwork}.  So long as both simulations sample the work distribution
adequately, they will have similar statistical error associated with them.  As
a cautionary note, low statistical errors can also be caused by inadequate
sampling  of the appropriate work distributions.  The statistical
error between two replicas can be reduced by decreasing the $\lambda$-distance
between them, and an optimal $\lambda$ schedule can reduce it for an entire
simulation.

\begin{table*}[htbp]
\begin{center}
\begin{tabular}{|l||c|c|}
\multicolumn{3}{c}{\bf (a) Acetamide Free Energy of Hydration: Current Work}\\ \hline 
& $\Delta F$ (kcal/mol) & $\epsilon_{RMS}$ (kcal/mol) \\
\hline 
% from /Users/matt/Documents/src/FES/ACE/ACE_22/*.out
% boostrap: make_bootstrap_summary('.', 'bootstrap_summary_ace_23', 'ACE
%     linear 21, 500/50 REX', 1900, 10000)
No Replica Exchange & -8.35 $\pm$ 0.051  & 0.120  \\ % ACE_22
Replica Exchange    & -8.14 $\pm$ 0.053  & 0.023  \\ % ACE_23
\hline
\multicolumn{3}{c}{}\\
\multicolumn{3}{c}{\bf (b) Acetamide Free Energy of Hydration: Literature}\\ \hline 
 & $\Delta G$ (kcal/mol) & Details \\  \hline
MacCallum and Tieleman \cite{maccallum:2003}& -8.25 $\pm$ 0.26 & TIP4P, TI\\
Shirts {\em et al.} \cite{shirts:2003b}&
   -8.20 \footnote{No long range van der Waals corrections} $\pm$ 0.03 &
TIP3P, TI\\
Chang {\em et al.} \cite{chang:2007}&    
   -8.54 $\pm$ 0.1 - 0.3 & TIP4P, BAR\\
Udier-Blagovi\'{c} {\em et al.} \cite{udier:2004}&    
   -9.65 $\pm$ 0.3 - 0.5 & TIP4P, FEP\\
\hline
Experimental \cite{wolfenden:1978}  & -9.54 & \\ 
\hline
\end{tabular}
\end{center}
\caption{ 
The hydration free energy of acetamide.  {\bf (a)} The Helmholtz hydration
free energy $\Delta F$ for the current work, as calculated by the Bennett
Acceptance Ratio, and the RMS hysteresis error. The $\Delta F$ statistical
errors are calculated by the bootstrap method.  {\bf (b)} Published values of
the Gibbs free energy $\Delta G$, obtained both computationally and
experimentally.  All computational results utilize the OPLS-AA force field
for the solute acetamide.
Also noted are the water model and free energy estimator (TI: Thermodynamic
Integration; FEP: Free Energy Perturbation; BAR: Bennett Acceptance Ratio)} 
% confirmed all literature numbers NPT
\label{tab:ACE}
\end{table*}

\subsection{Hysteresis Error and Replica Exchange}
\begin{figure*}
\begin{center}
\includegraphics[width=6.25in]{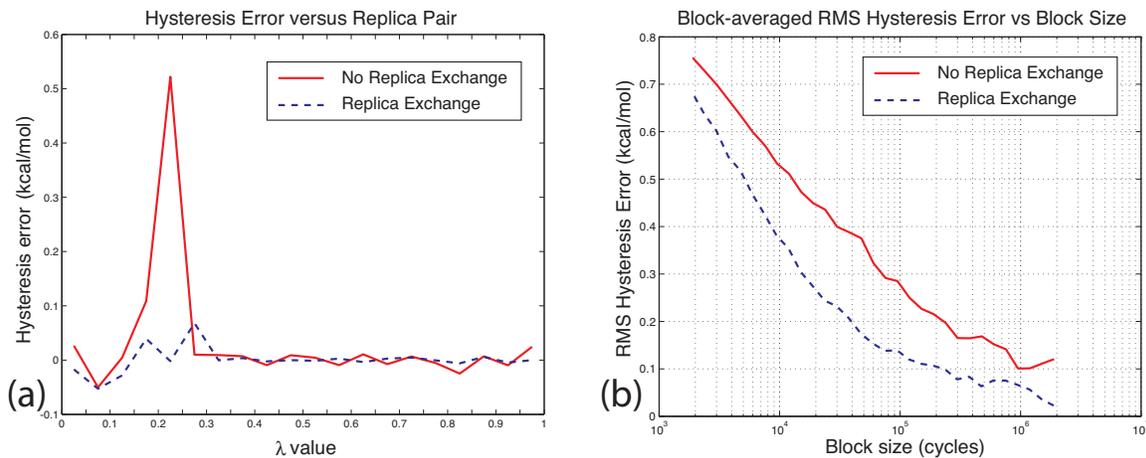}
\end{center}
\caption{
{\bf (a)} The hysteresis error between neighboring replicas.  Replica exchange
effectively reduces the hysteresis error for replica pairs.  {\bf (b)} Block
averages of the RMS hysteresis error, showing that the hysteresis error falls
with increasing block size.  Replica exchange increases the rate at which
hysteresis error is lowered, thereby achieving the same magnitude error with
simulations which are on average 4-8 times shorter.  }
\label{fig:ACE_hyst_error} 
\end{figure*}

For a fixed $\lambda$ schedule, the hysteresis error may be reduced with
either an improved sampling methodology like replica exchange, or longer
simulations per replica.  The effects of both approaches are illustrated in
Fig. \ref{fig:ACE_hyst_error}.  

Panel (a) shows $\epsilon_H$ for each neighboring replica pair.  The
hysteresis error is not uniform across all pairs, with spikes in the region
$\lambda=0.1 - 0.3$.  Replica exchange systematically reduces the hysteresis
error for all pairs of replicas.

Panel (b) illustrates how both longer sampling and replica exchange affect the
hysteresis error.  Block averaging shows that the average RMS hysteresis error
declines consistently with longer simulations.  This reduction can be improved
with replica exchange; in fact, a simulation with replica exchange will
achieve the same magnitude of RMS hysteresis error 5 times more quickly than
one without replica exchange.

\subsection{Average Swap Probability}
\begin{figure}[htbp]
\begin{center}
\includegraphics[width=3.25in]{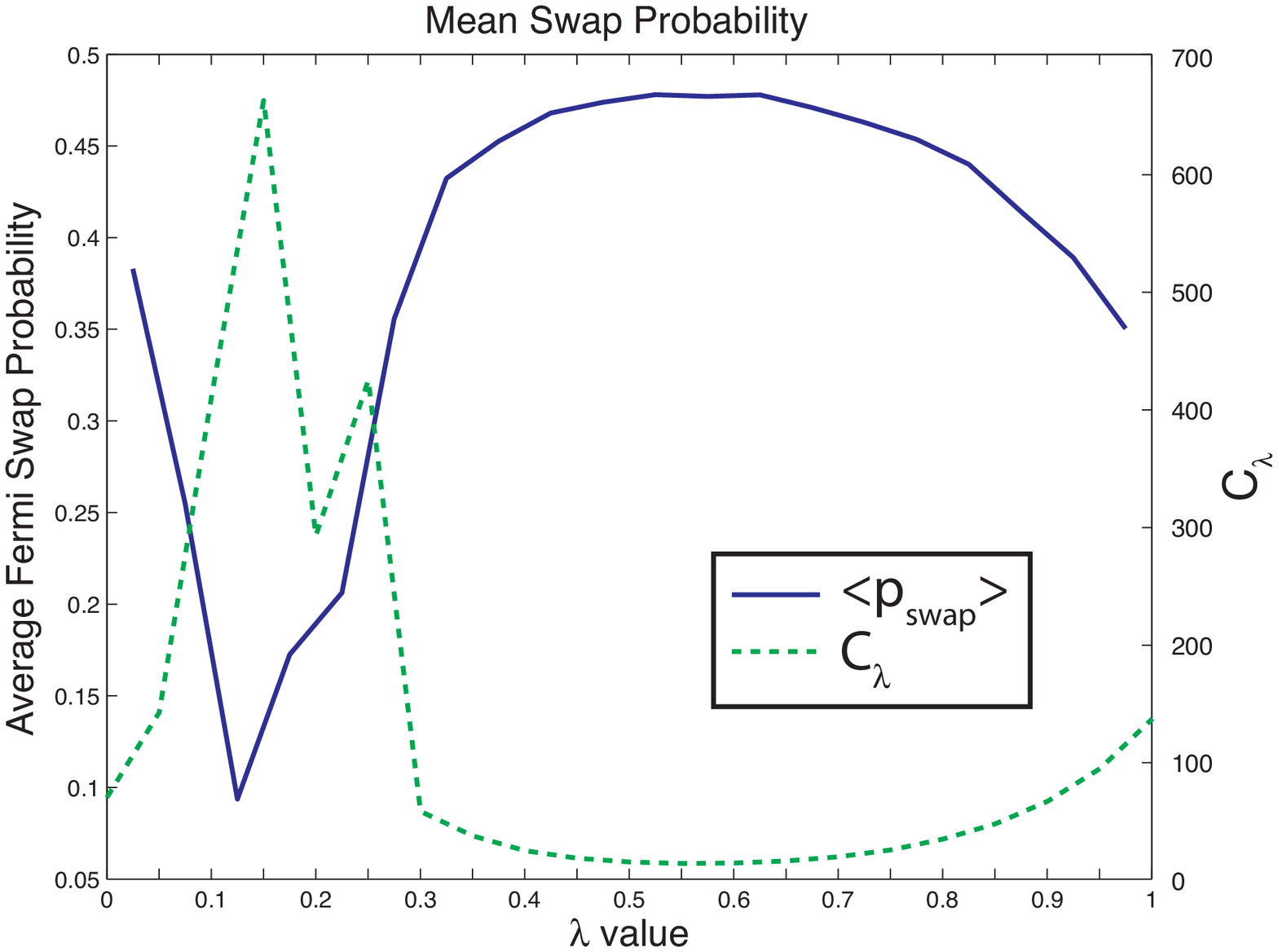}
\end{center}
\caption{The average swap probability between adjacent replicas and
$C_\lambda$=var($\partial U/\partial \lambda$) evaluated for each replica
(from the replica exchange simulation; simulation with no replica
exchange is not significantly different).  Spikes in $C_\lambda$ indicate
regions of low swap probability.  }
\label{fig:ACE_swap_prob} 
\end{figure}

Fig. \ref{fig:ACE_swap_prob} shows downward spikes in the swap probability for
values of $\lambda$ where the hysteresis error is large in Fig.
\ref{fig:ACE_hyst_error}(a).  These results are consistent with the proposal
that swap probability between two replicas is an indicator of the rate at
which $\epsilon_H$ is minimized.  The same region is characterized by a
positive spike in $C_\lambda$, which is expected based on the relationship
between the swap probability and $C_\lambda$ in Eq. \eqref{eqn:pswap_lin}.
However, while the swap probability calculation requires two separate
simulations, estimates of $C_\lambda$ can be obtained from just one.
Moreover, $\langle p_{swap}\rangle$ varies as the distance between the
replicas changes, complicating the interpretation if the $\lambda$ schedule is
not uniform.  Evaluation of  $C_\lambda$ as a function of $\lambda$ using a
preliminary, coarse $\lambda$ schedule can identify regions where the swap
probability is expected to be low, and can be used to construct optimal
$\lambda$ schedules, as discussed in Sec. \ref{sec:optimal}.

%%%%%%%%%%%%%%%%%%%%%%%%%%%%%%%%%%%%%%%%%%%%%%%%%%%%%%%%%%%%%%%%%%%%%%%%%%%%%%%

\section{Discussion}

\subsection{Physical Interpretation of $C_\lambda$ Profile}

\begin{figure}[htbp]
\begin{center}
\includegraphics[width=3.25in]{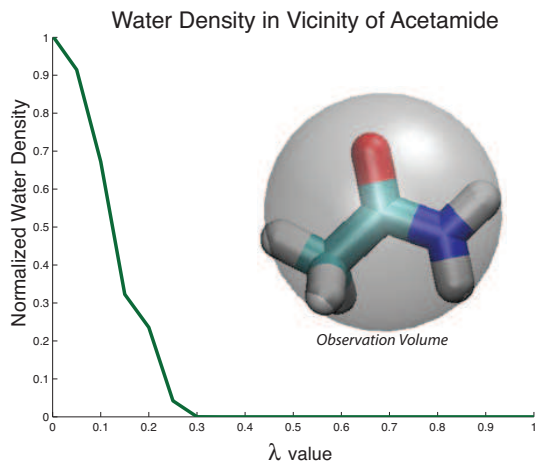}
\end{center}
\caption{
Water density within 2.5\AA{} of the acetamide carbonyl carbon as $\lambda$
varies.  The inset illustrates the position and size of the observation
volume with respect to an ACE molecule.  Density is normalized by the bulk
density.  As $\lambda$ increases, waters are expelled by the growing cavity.
 }
\label{fig:water_occupancy} 
\end{figure}

To gain a physical interpretation of the profile for $C_\lambda$ shown in Fig.
\ref{fig:ACE_swap_prob}, we plot in Fig.  \ref{fig:water_occupancy} the
average water density in a 2.5\AA{} sphere surrounding the carbonyl carbon of
acetamide.  The plot shows that water occupancy around the growing solute
decreases rapidly in the range of $\lambda\sim0.15$.  The expulsion and
rearrangement of water molecules during cavitation leads to a large shift in
the equilibrium ensemble, giving rise to a pronounced spike in $C_\lambda$.
(Smaller shifts in $C_\lambda$ near $\lambda=1$ reflect electrostatic effects
and are not observed for simulations where $\lambda_C=0$, data not shown.)
Thus, $C_\lambda$ profiles serve as useful probes for detecting large shifts
in equilibrium ensembles.  Regions where the equilibrium ensembles change most
rapidly are the regions that contribute to the largest errors in free energy
calculations.

\subsection{Optimal $\lambda$ Schedule for Free Energy
Calculations}\label{sec:optimal}
For given computational resources, with the number of replicas and the
simulation length fixed,  the RMS hysteresis error of a simulation may be
decreased by optimizing the $\lambda$ schedule, or the distribution of
$\lambda$ across the replicas.  The swap probability gives the rate at which
the average hysteresis error falls between two replicas, and in an optimized
simulation it would be uniform across all replica pairs.  In practice it is
difficult to obtain the $\lambda$ schedule which makes the swap probability
exactly uniform, but reasonable approximations can be made by using the
linearized swap probability, given by Eq. \eqref{eqn:pswap_lin}.

First, it is necessary to perform some number of preliminary simulations to
obtain $C_\lambda$ along a coarse $\lambda$ schedule.  These initial
simulations need not be as long as the final production runs, since
$C_\lambda$ converges more quickly than $\delta F$ and is more tolerant of
error.  With a rough estimate of $C_\lambda(\lambda)$ in hand, the $\lambda$
schedule can be adjusted to ensure that the linear swap probability is uniform
between all replicas.  Alternatively, one might simply shift replicas from
where $C_\lambda$ is small to where it is large.  Both approaches are only
approximate, and break down when the linear response assumption in Eq.
\eqref{eqn:pswap_lin} ceases to be valid.  They may be applied iteratively as
$C_\lambda$ is evaluated for new $\lambda$ schedules.

The aim of an optimal $\lambda$ schedule is to place replicas close together
in regions where the  $C_\lambda$ profile shows spikes.  This ensures
reasonable swap probabilities and minimal hysteresis errors in regions that
are problematic.  Preliminary investigations show that even when the schedule
is improved in an {\em ad hoc} manner, hysteresis as well as statistical
errors decrease.  

\subsection{Replica Exchange}

Replica exchange provides a Monte Carlo move which may allow a replica to
access a distant part of its equilibrium ensemble in one step.  It is no
substitute for conformational exploration within a replica.  This point, while
obvious, must be emphasized in the context of the hysteresis error, which does
not report on the quality of intra-replica sampling.  As an extreme but
illustrative case, consider a system of some number of frozen replicas, each
with a different configuration, which undergo replica exchange moves but no
conformational changes.  With just a modest number of swaps, these
configurations attain the probability distribution described by Eq.
\eqref{eqn:db}, and the hysteresis error is zero.  The system has achieved
inter-replica equilibrium, but the intra-replica probability distribution has
not been obtained.  In practice, the majority of Monte Carlo moves must be
within a replica.  The optimal frequency of swap moves remains an open
question, although preliminary simulations suggest that more frequent swaps
reduce the hysteresis error more quickly.

\section{Summary and Conclusion}

In a simulation of multiple replicas, each sampling the equilibrium ensemble
of a different Hamiltonian, swapping configurations between replicas is a
nonequilibrium work process.  Accordingly, the work needed to perform such
swaps has a distribution of values, as described by the Crooks fluctuation
theorem.  The hysteresis error $\epsilon_H$ developed here measures how
closely a given simulation reproduces these work distributions between a pair
of replicas.

The hysteresis error is particularly useful in the context of free energy
calculations.  It reports on the combined bias of the forward and reverse free
energy perturbation techniques, and it measures how completely individual
replicas sample their equilibrium ensemble.  The RMS hysteresis error, which
reports on $\epsilon_H$ for the whole $\lambda$ schedule, may be decreased by
running a longer simulation, employing replica exchange,  utilizing an
improved $\lambda$ schedule, or all of these approaches.

The average swap probability is another useful measure and can be calculated
whether or not replica exchange is employed.  Since it determines the rate at
which the hysteresis error decreases with simulation length, the swap
probability can be used to optimize the $\lambda$ schedule.  With a uniform
average swap probability the hysteresis error falls evenly between all replica
pairs.  This maximizes the efficiency of simulations with fixed computational
resources, avoiding unnecessary replicas where the hysteresis is low and
preventing excessive errors from regions where the hysteresis error is large.

Furthermore, the swap probability, along with a related measure $C_\lambda$,
yields insight into the microscopic behavior of a system.  The swap
probability is low and $C_\lambda$ is large when the equilibrium ensemble 
changes rapidly with $\lambda$.  Slow convergence and bias errors in free
energy calculations arise when there are spikes in the $C_\lambda$ profile
along the $\lambda$ schedule, which coincides with large hysteresis errors.

\begin{acknowledgments}
We are grateful to Hoang Tran, Albert Mao, Xiaoling Wang, Nicholas Lyle, and
in particular Alan Chen and Andreas Vitalis, for helpful discussions and
insightful comments.  This work was supported by grants MCB - 0416766 and MCB
- 0718924 from the National Science Foundation.
\end{acknowledgments}

\appendix 
%%%%%%%%%%%%%%%%%%%%%%%%%%%%%%%%%%%%%%%%%%%%%%%%%%%%%%%%%%%%%%%%%%%%%%%%%%%%%%%

\section{Derivations}
\subsection{Fluctuation theorem derivation}\label{sec:ft}
% from notes, 10/19/07.1
We derive the Crooks fluctuation theorem \eqref{eqn:Pwork} in the context of
instantaneously switching $\lambda_0 \rightarrow \lambda_1$ (forward) and
$\lambda_1 \rightarrow \lambda_0$ (reverse). Expanding the ratio $\rho_0 /
\rho_1$ with \eqref{eqn:rho_FU} for an arbitrary configuration $\Gamma$,
\begin{subequations}
\begin{eqnarray}
\frac{\rho_0(\Gamma)}{\rho_1(\Gamma)} &=&
   \exp \left[ \beta(F_0 - F_1) - \beta(U_0 - U_1)\right],\nonumber\\
&=& \exp(-\beta \delta F+ \beta W^F),\nonumber \\
&=& \exp(\beta W_D^F)\label{eqn:r0r1},
\end{eqnarray}
and similarly,
\begin{equation}
\frac{\rho_1(\Gamma)}{\rho_0(\Gamma)} = \exp[\beta W_D^R(\Gamma)] 
\label{eqn:r1r0}
\end{equation}
\end{subequations}
where the definitions of work \eqref{eqn:work} and dissipated work
\eqref{eqn:work_d} were used.

We integrate $\rho_1$ from \eqref{eqn:r0r1} over all configurations, but
consider contributions only from those $\Gamma$ for which the forward
dissipated work work value takes on a specific value, $W_D$:
\begin{eqnarray}
\int d\Gamma \rho_0(\Gamma) \exp[-\beta W_D^F(\Gamma)] \delta[\beta W_D -
\beta W_D^F(\Gamma)] \nonumber \\
= \int d\Gamma \rho_1(\Gamma) \delta[\beta W_D - \beta
W_D^F(\Gamma)]\label{eqn:dwd1}.
\end{eqnarray}
Since, from \eqref{eqn:r0r1} and \eqref{eqn:r1r0},
$$ W_D^F(\Gamma) = -W_D^R(\Gamma)$$
\eqref{eqn:dwd1} becomes,
\begin{eqnarray}
\int d\Gamma \rho_0(\Gamma) \exp[-\beta W_D^F(\Gamma)] \delta[\beta W_D -
\beta W_D^F(\Gamma)] \nonumber \\
= \int d\Gamma \rho_1(\Gamma) \delta[\beta W_D + \beta W_D^R(\Gamma)].
\label{eqn:dwd2}
\end{eqnarray}

We define $P^F(W_D)$ as the probability of observing a given dissipated work
value in the forward switching process, and it can be expressed as an integral
over all configurations which yield this value,
\begin{subequations}
\begin{equation}
P^F(W_D) = \int d\Gamma \rho_0(\Gamma) \delta[\beta W_D - \beta W_D^F(\Gamma)]
\end{equation}
Likewise, the probability of observing a given disspated work value in the
reverse switching process is,
\begin{equation}
P^R(W_D) = \int d\Gamma \rho_1(\Gamma) \delta[\beta W_D - \beta W_D^R(\Gamma)]
\end{equation}
\end{subequations}
With these definitions, \eqref{eqn:dwd2} may be written as,
$$ \exp(-\beta W_D) P^F(\beta W_D) = P^R(-\beta W_D),$$
which is equivalent to \eqref{eqn:Pwork}.

\subsection{Fluctuation theorem and hysteresis error}\label{sec:hysteresis_error}
%from notes, 10/1/07
The relationship between some arbitrary deviation of a simulation from the
Crooks fluctuation theorem and the hysteresis error is derived by first
rewriting Eq. \eqref{eqn:FT_error} as,
\begin{equation}
  P_R^\ast(-\beta W_D) \exp(\beta W_D) =
P_F^\ast(\beta W_D) \exp(-\beta \epsilon_{FT}^\ast). \label{eqn:prpf}
\end{equation}
Inserting the $\delta F_{FEP}^R$ definition \eqref{eqn:fep_R} into the
definition of the hysteresis error \eqref{eqn:hyst_def}, expanding the reverse
work with \eqref{eqn:wdr} and using the $\delta F_{FEP}^F$ estimate for
$\delta F$,
\begin{eqnarray} 
\epsilon_H &=& \delta F_{FEP}^F - \beta^{-1}\ln\langle\exp(-\beta
W^R)\rangle_1^\ast \nonumber,\\
&=& \delta F_{FEP}^F -\beta^{-1}\ln \left [ \langle\exp(-\beta W^R_D)\rangle_1^\ast
 \exp(\beta \delta F_{FEP}^F) \right]\nonumber, \\
&=& -\beta^{-1}\ln \left [ \langle\exp(-\beta W^R_D)\rangle_1^\ast
 \right]\nonumber. 
\end{eqnarray}
We now expand the estimated ensemble average as an integral over all values of
$\beta W_D^F$, with $P_R^\ast$ the normalized histogram of $\beta W_D^R$
obtained from a simulation,
\begin{equation*}
\epsilon_H = -\beta^{-1} \ln \left[ \int_{-\infty}^{+\infty} d[\beta W_D^R]
P_R^\ast(\beta W_D^R)\exp(-\beta W_D^R) \right].
\end{equation*}
As $\beta W_D^R$ is a dummy variable, we  change it to $-\beta W_D$,
\begin{equation*}
\epsilon_H = -\beta^{-1} \ln \left[ \int_{-\infty}^{+\infty} d[\beta W_D]
P_R^\ast(-\beta W_D)\exp(\beta W_D) \right],
\end{equation*}
where we implicitly multiplied the integrand by $-1$ to preserve the limits of
integration.  With \eqref{eqn:prpf} the above can be written as,
\begin{equation*}
\epsilon_H= -\beta^{-1} \ln \left[ \int_{-\infty}^{+\infty} d(\beta W_D)
P_F^\ast(\beta W_D) \exp(-\beta \epsilon_{FT}^\ast) \right],
\end{equation*}
which reduces to \eqref{eqn:hyst_error_FT}.  

\subsection{Inter-replica equilibrium and hysteresis error}\label{sec:rho_e}
We can relate an small arbitrary error in the calculated distribution
$\rho'_N$ to the hysteresis error by considering a small error
$\rho_\epsilon(\Gamma_0, \Gamma_1)$ in the otherwise correctly estimated
$\rho'_N$.  Rewriting \eqref{eqn:db},
\begin{equation*}
\rho'_N + \rho_\epsilon = \rho_N \exp(-\beta \Delta U_{swap}),
\end{equation*}
we integrate over all configuration pairs and rewrite $\Delta U_{swap}$ with
\eqref{eqn:deltaU_W},
\begin{eqnarray}
\int d\Gamma_0 d\Gamma_1 \rho'_N &+& \int d\Gamma_0 d\Gamma_1  \rho_\epsilon
\nonumber \\
&=& \int d\Gamma_0 d\Gamma_1 \rho_0(\Gamma_0) \rho_1(\Gamma_1) \exp[-\beta W^F(\Gamma_0)]
\nonumber\\
&\times& \int d\Gamma_1 \rho_1(\Gamma_1) \exp[-\beta W^R(\Gamma_1)]\label{eqn:c1}.
\end{eqnarray}
With the sampling error contained in $\rho_\epsilon$, the $\rho'_N$ term
(expanded with \eqref{eqn:rnp_def}) is identically one.
Taking the logarithm and dividing by $\beta$, \eqref{eqn:c1}
becomes,
\begin{equation}
-\beta^{-1} \ln \left[ 1 + \int d\Gamma_0 d\Gamma_1 \rho_\epsilon \right] = 
\delta F_{FEP}^R - \delta F_{FEP}^F\label{eqn:c2},
\end{equation}
where we have used the $\delta F_{FEP}$ definitions \eqref{eqn:fep}.
With the approximation $\ln(1+x) \simeq x$ for small $x$ and the definition of
$\epsilon_H$ \eqref{eqn:hyst_def}, we obtain Eq.
\eqref{eqn:epsilon_h_rho_S}.

\subsection{Linearized average swap probability} \label{sec:pswap}
% See notes 3/23/07 and 3/26/07
Here we consider the average Fermi swap probability between two replicas whose
$\lambda$ parameters differ by a small amount, $\delta$ (written as
$\delta_\lambda$ in the text).  For convenience we define 
\begin{eqnarray*}
\mu &\equiv& \beta \Delta U_{swap},\\
&=& \beta[U_\delta(\Gamma_0) - U_0(\Gamma_0) + U_0(\Gamma_\delta) -
U_\delta(\Gamma_\delta)],
\end{eqnarray*}
where $\Gamma_0$ and $\Gamma_\delta$ are configurations drawn from the
equilibrium distributions $U_0$ and $U_\delta$ parameterized by $\lambda_0$
and $\lambda_0 + \delta$, respectively.  We expand $U_\delta$ as a
Taylor series about $\lambda_0$,
$$ U_\delta(\Gamma) = U_0(\Gamma) + \delta V_0(\Gamma) +
\frac{\delta^2}{2}W_0(\Gamma) + O(\delta^3),$$
with
\begin{eqnarray*}
V_0 &\equiv& \left. \frac{\partial U}{\partial \lambda} \right|_{\lambda =
\lambda_0}, \\
W_0 &\equiv& \left. \frac{\partial^2 U}{\partial \lambda^2} \right|_{\lambda =
\lambda_0}.
\end{eqnarray*}
$\mu$ can then be written as,
$$\mu = \beta \delta [V_0(\Gamma_0) - V_0(\Gamma_\delta)] + 
\frac{\beta \delta^2}{2} [W_0(\Gamma_0) - W_0(\Gamma_\delta)].$$
Note that $\mu$ is small ($O(\delta)$); thus, with the identities,
\begin{subequations}
\begin{eqnarray}
\exp(x) &=& 1 + x + x^2 / 2 + ... ,\\
\frac{1}{1+x} &=& 1 - x + x^2 -... ,\label{eqn:frac}
\end{eqnarray}
\end{subequations}
we may write the Fermi swap probability between configurations $\Gamma_0$ and
$\Gamma_\delta$ as,
\begin{eqnarray*}
p_{swap} &=& \frac{1}{1 + \exp{\mu}},\nonumber\\
&=& \frac{1}{2} \left( \frac{1}{1 + \mu/2 + \mu^2/4 + O(\mu^3)} \right),
\nonumber\\
&=& \frac{1}{2} \left[ 1 - (\mu/2 + \mu^2/4) + (\mu/2 +
\mu^2/4)^2 + O(\mu^3)\right],\nonumber\\
&=& \frac{1}{2} - \frac{1}{4} \mu + O(\mu^3).
\end{eqnarray*}

The average swap probability is the ensemble average over all configuration
pairs,
\begin{eqnarray}
\langle \langle p_{swap}\rangle_0 \rangle_\delta &=& 
\frac{1}{2} - \frac{1}{4}\langle \langle \mu \rangle_0 \rangle_\delta,
\nonumber\\
&=& \frac{1}{2} - \frac{1}{4}\left( \beta \delta \langle V_0 \rangle_0 +
\frac{\beta\delta^2}{2}\langle W_0 \rangle_0 \right.\nonumber\\
&&\left. - \beta \delta \langle
V_0\rangle_\delta -\frac{\beta \delta^2}{2} \langle W_0 \rangle_\delta
\right) \nonumber\\
&&+ O(\delta^3)\label{eqn:ps1}.
\end{eqnarray}
To evaluate $\langle \cdot \rangle_\delta$, we first obtain $Q_\delta$,
the partition function at $(\lambda_0 + \delta)$:
\begin{eqnarray*}
Q_\delta &\equiv& \int d\Gamma \exp(-\beta U_\delta), \\
&=& \int d\Gamma \exp(-\beta U_0) \left [1 - \beta \delta V_0 + 
O(\delta^2)\right],\\
&=& Q_0 \left[1 - \beta \delta \langle V_0\rangle_0  + O(\delta^2)\right],
\end{eqnarray*}
and its reciprocal,
\begin{equation*}
Q_\delta^{-1} = Q_0^{-1}\left[1 + \beta \delta \langle V_0 \rangle_0  +
O(\delta^2)\right].
\end{equation*}

We can now evaluate $\langle V_0\rangle_\delta$ and $\langle
W_0\rangle_\delta$, retaining only terms which will remain $O(\delta^2)$ or
larger in \eqref{eqn:ps1}:
\begin{eqnarray*}
\langle V_0\rangle_\delta &\equiv& 
Q_\delta^{-1} \int d\Gamma \exp(-\beta U_\delta) V_0, \\
&=& Q_0^{-1} \left( 1+\beta \delta \langle V_0\rangle_0\right) 
   \int d\Gamma (1-\beta \delta V_0) \exp(-\beta U_0) V_0, \\
&=& (1+\beta \delta \langle V_0\rangle_0)(\langle V_0\rangle_0 - \beta \delta
   \langle V_0^2 \rangle_0), \\
&=& \langle V_0 \rangle_0 + \beta \delta \left( \langle V_0 \rangle_0^2 -
   \langle V_0^2 \rangle_0\right),
\end{eqnarray*}
and
\begin{eqnarray*}
\langle W_0\rangle_\delta &\equiv& Q_\delta^{-1} \int d\Gamma \exp(-\beta
U_\delta) W_0 ,\\
&=& Q_0^{-1}(1 + O(\delta)) \int d\Gamma \exp(-\beta U_0) W_0 [1 - O(\delta)],\\
&=& \langle W_0\rangle_0 + O(\delta).
\end{eqnarray*}
Finally, \eqref{eqn:ps1} becomes,
\begin{equation}
\langle \langle p_{swap}\rangle_0 \rangle_\delta = \frac{1}{2} - \frac{\beta^2
\delta^2}{4}\left( \langle V_0^2\rangle_0 - \langle V_0\rangle^2_0\right) +
O(\delta^3),
\end{equation}
equivalent to Eq. \eqref{eqn:pswap_lin}, which is valid for small $\delta$.

%%%%%%%%%%%%%%%%%%%%%%%%%%%%%%

\section{$U_{LJ}$ and $U_C$ functional forms\label{sec:scaling}}
The functional forms of both the Coulomb and Lennard-Jones potentials were
developed for this work based on three criteria: 
\begin{enumerate} 
\item Configurations where the solute and solvent overlap may be observed for
$\lambda=0$.  For such configurations, we require:
\begin{itemize}
\item That swaps be permitted with reasonable frequency for small $\lambda$
(e.g.  $\lambda = 0.1$).
\item That swap probabilities falls off quickly thereafter; in particular, we
wish to avoid the situation where the swap probability declines only very near
$\lambda=1.0$.
\end{itemize}

\item We require that $\partial U / \partial \lambda$ is not always zero for
$\lambda=0$ to avoid complications with the Thermodynamic Integration (TI)
estimator.  While, we do not report results using TI in this work, 
we wish to construct a $\lambda$ schedule that works with all estimators.

\item In this work, $\lambda_{LJ}=\lambda_C$.  Therefore, Lennard-Jones
repulsion must dominate Coulombic attraction at very small atomic separations.
\end{enumerate}
While various ways to scale the potential have been discussed in the
literature \cite{beutler:1994, pitera:2002, shirts:2005a}, none of these
satisfied all of our requirements.  It should be noted that condition 3 is
somewhat arbitrary, and more common scaled potentials may be used if the
insertion process scales the Lennard-Jones prior to the Coulomb potential.

\paragraph{Coulomb scaling}
We employ a modified version of the linear soft-core
scaling \cite{pitera:2002};  for two atoms of charges $q_i$ and $q_j$ distance
$r$ apart, the potential energy is 
$\lambda_C$ as,
\begin{equation}
U_C(r,\lambda_C) = \lambda_C \frac{q_i q_j}{\alpha_C (1-\lambda_C) + r},
\end{equation}
$\alpha_C$ controls the ``soft core'' term, and for small $\lambda_C$ imposes a
minimum effective atomic separation.  $\alpha_C = 1.5\AA{}$ for all
simulations in this work.

\paragraph{Lennard-Jones scaling}
The Lennard-Jones potential between two particles may be written generally as,
\begin{equation}
U_{LJ}(r,\lambda_{LJ}) = B A (A - 1),
\label{eqn:U_LJ}
\end{equation}
where, for unscaled Lennard-Jones,
\begin{equation*}
A(r) = \left( \frac{\sigma}{r} \right)^6,  ~~~~~~~~~
B = 4 \epsilon.
\end{equation*}

Simple linear scaling by $\lambda_{LJ}$ of the Lennard-Jones potential is
known to be unsatisfactory, and a number of alternate forms have been
introduced.  We developed the exponential soft-core,
\begin{subequations} %cf amsldoc.pdf section 3.11.3
\label{eqn:exp_sc}   %cf http://www.iam.ubc.ca/~newbury/tex/subeqn.html
\begin{eqnarray}
A(r,\lambda_{LJ}) &=& 1/ \left [\alpha_{LJ}(1-\lambda_{LJ})^b + 
   \left( \frac{r}{\sigma} \right)^6 \right],\label{eqn:exp_sc_A} \\
B(\lambda_{LJ}) &=& 4 \epsilon \frac{1-e^{-k\lambda_{LJ}}}{1-e^{-k}},
   \label{eqn:exp_sc_B}
\end{eqnarray}
\end{subequations}
with $a=4$, $k=1$ and $\alpha_{LJ} = 0.5\AA{}$.  The precise position along
the $\lambda$ coordinate of the swap probability trough (see Fig.
\ref{fig:ACE_swap_prob}) is specific to this Lennard-Jones potential.

\bibliography{MAW-RVP} 
\end{document}